\documentclass[prb,,twocolumn]{revtex4-1} 

\usepackage{amsmath}  
\usepackage{amsfonts} 
\usepackage{color}
\usepackage{verbatim}
\usepackage[pdftex]{graphicx}
\usepackage{epstopdf} 
\usepackage{hyperref}

\begin{document}


\title{Using split-ring resonators to measure the electromagnetic properties of materials:\\ An experiment for senior physics undergraduates}

\author{J. S. Bobowski}
\email{jake.bobowski@ubc.ca} 
\altaffiliation[permanent address: ]{3333 University Way, 
  Kelowna, Canada} 
\affiliation{Department of Physics, University of British Columbia Okanagan, Kelowna, BC V1V 1V7}


\date{\today}

\begin{abstract}
A spilt-ring resonator experiment suitable for senior physics undergraduates is described and demonstrated in detail.  The apparatus consists of a conducting hollow cylinder with a narrow slit along its length and can be accurately modelled as a series $LRC$ circuit.  The resonance frequency and quality factor of the split-ring resonator are measured when the apparatus is suspended in air, submerged in water, and submerged in an aqueous solution of various concentrations of NaCl.  The experimental results are used to extract the dielectric constant of water and to investigate the dependence of the resonator quality factor on the conductivity of the NaCl solution.  The apparatus provides opportunities to experimentally examine radiative losses, complex permittivity, the electromagnetic skin depth, and cutoff frequencies of rf propagation in cylindrical waveguides, which are all concepts introduced in an undergraduate course in electrodynamics.  To connect with current research, the use of split-ring resonators as a tool to precisely measure the electromagnetic properties of materials is emphasized.

\end{abstract}

\maketitle 

\section{Introduction} 

The split-ring resonator (SRR) is currently used in two areas of active of physics research.  Two-dimensional periodic arrays of SRRs are key components of so-called metamaterials that can have a permittivity and permeability that are simultaneously negative at microwave frequencies.\cite{Smith:2000, Smith:2001}  The second major use of SRRs is as an apparatus to make precision measurements of the electromagnetic (EM) properties of a material of interest.  In one remarkable study, a nanoscale SRR probe was used to directly measure the magnetic fields of light at optical frequencies.\cite{Burresi:2009}  As another example, high-$Q$ SRRs are used to measure the temperature dependence of the magnetic penetration depth and the surface resistance of superconducting samples with sub-angstrom  and micro-ohm resolution respectively.\cite{Hardy:1993, Bobowski:2010, Bonn:1991} The primary purpose of the experiment described here is to use the SRR as a tool to study the EM properties of aqueous solutions of NaCl.  Careful characterization of the complex permittivity of materials continues to be an important and active area of research.\cite{Bobowski:2012a, Bobowski:2012b} It is also worth noting that numerous applications, such as bandpass and bandstop filters, that make use of the EM properties of SRRs are actively being developed.\cite{Ricci:2006, Xiao:2007} 

As a teaching tool, a SRR experiment offers numerous appealing features.  The lumped circuit element model facilitates a theoretical treatment that uses methods familiar to students, yet requires special insights that go sufficiently beyond standard series $LRC$ results such that the analysis remains interesting and engaging.  The experimental apparatus is relatively simple enabling students to fully appreciate the techniques used to acquire the data.  The experiment also allows students to explore, in a laboratory setting, concepts common to all undergraduate physics curricula such as radiative losses, complex permittivities, the skin depth of conductors, cutoff frequencies in cylindrical waveguides, the reduction of wavelength in high-permittivity materials, and, in principle, the effect of high-permeability materials on both the resonance frequency and quality factor of the SRR.  Finally, as a practical matter, the equipment required for the experiment is either relatively easy and inexpensive to acquire or is already existing in most university or college physics departments.  

The outline of this paper is as follows: Section~\ref{sec:II} considers both theoretically and experimentally the resonance frequency $f_0$ and quality factor $Q$ of the SRR while suspended in air.  Due to radiative losses, the measured $Q$ is significantly lower than the value predicted using the simple $LRC$ model.  The radiative losses are suppressed by suspending the SRR inside of a conducting cylindrical waveguide of suitable diameter.  In Sec.~\ref{sec:III}, the cylindrical waveguide is filled with water.  The water fills the gap of the SRR thereby modifying the capacitance and hence $f_0$ and $Q$.  The shift in $f_0$ is used to determine the real part of the relative permittivity $\varepsilon^\prime$ of water.  The observed change in $Q$ implies additional losses that arise from the imaginary part of the relative permittivity $\varepsilon^{\prime\!\prime}$.  Next, the water is made conducting by adding known amounts of NaCl.  For low concentrations of NaCl, the resonance frequency is unaffected.  However, the conductivity of the water, denoted $\sigma$, adds an additional loss mechanism which further degrades the $Q$. Section~\ref{sec:IV} investigates the dependence of $Q$ on $\sigma$.  Section~\ref{sec:V} summarizes the key results.  Throughout, it is assumed that the reader is familiar with standard $LRC$ resonators.  Derivations presented in the following sections make use the lumped circuit element model and are presented in a way that is accessible to undergraduates students familiar with basic electronics and complex algebra.  Finally, a list of the all of the equipment and materials required is given in the appendix.  Where appropriate, possible vendors are suggested and cost estimates are made.

\section{Suspended in air}\label{sec:II}
\begin{figure}[t] 
\centering
\begin{tabular}{cc}
(a)\includegraphics[width=3.9 cm]{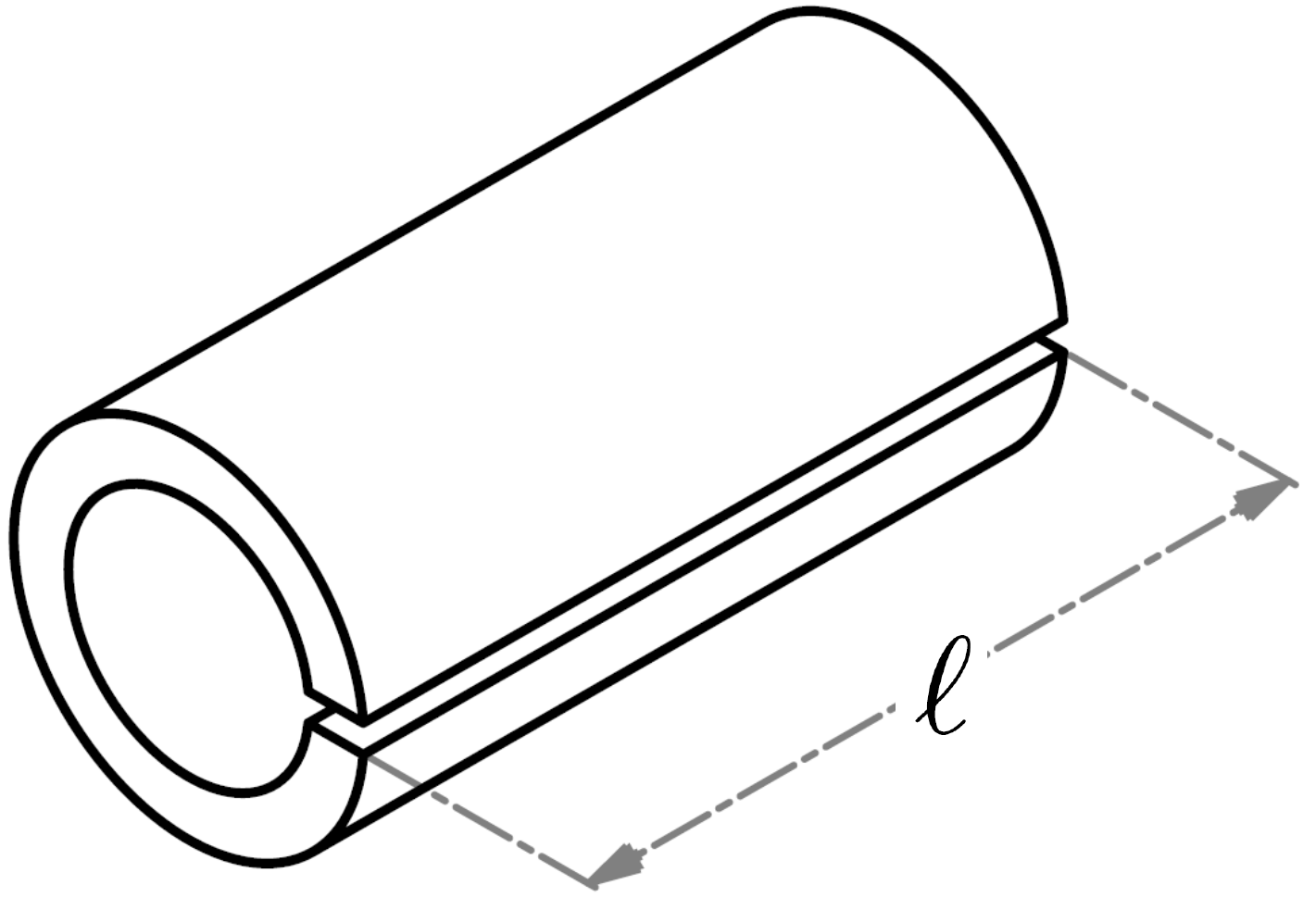} & ~~~(b)\includegraphics[width=2.4 cm]{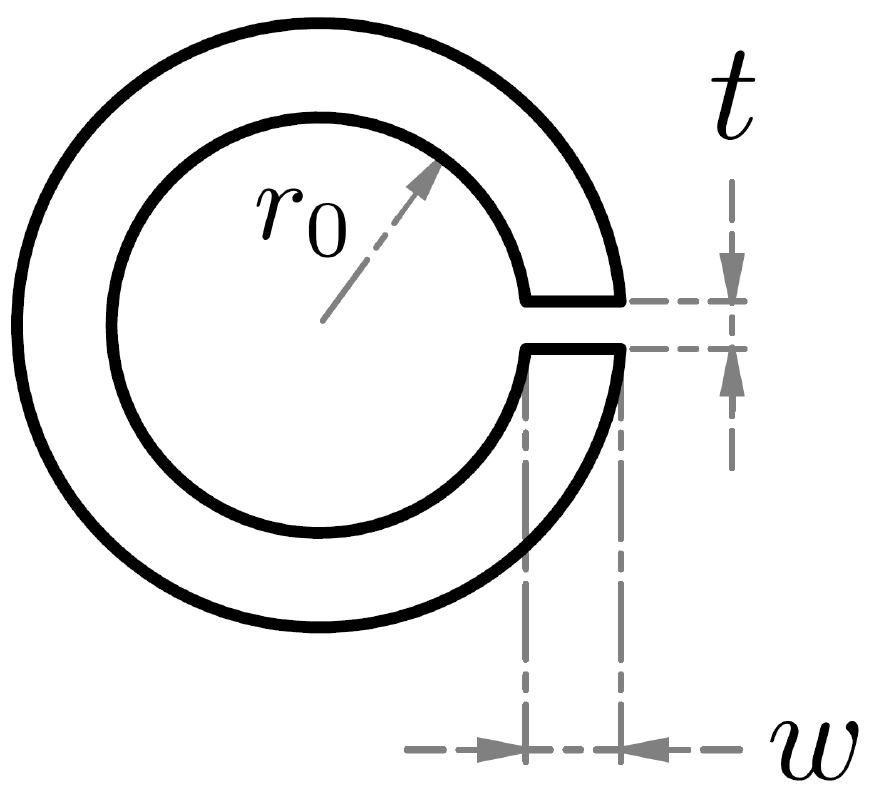}\\
~ & ~\\ 
(c)\includegraphics[width=4.7 cm]{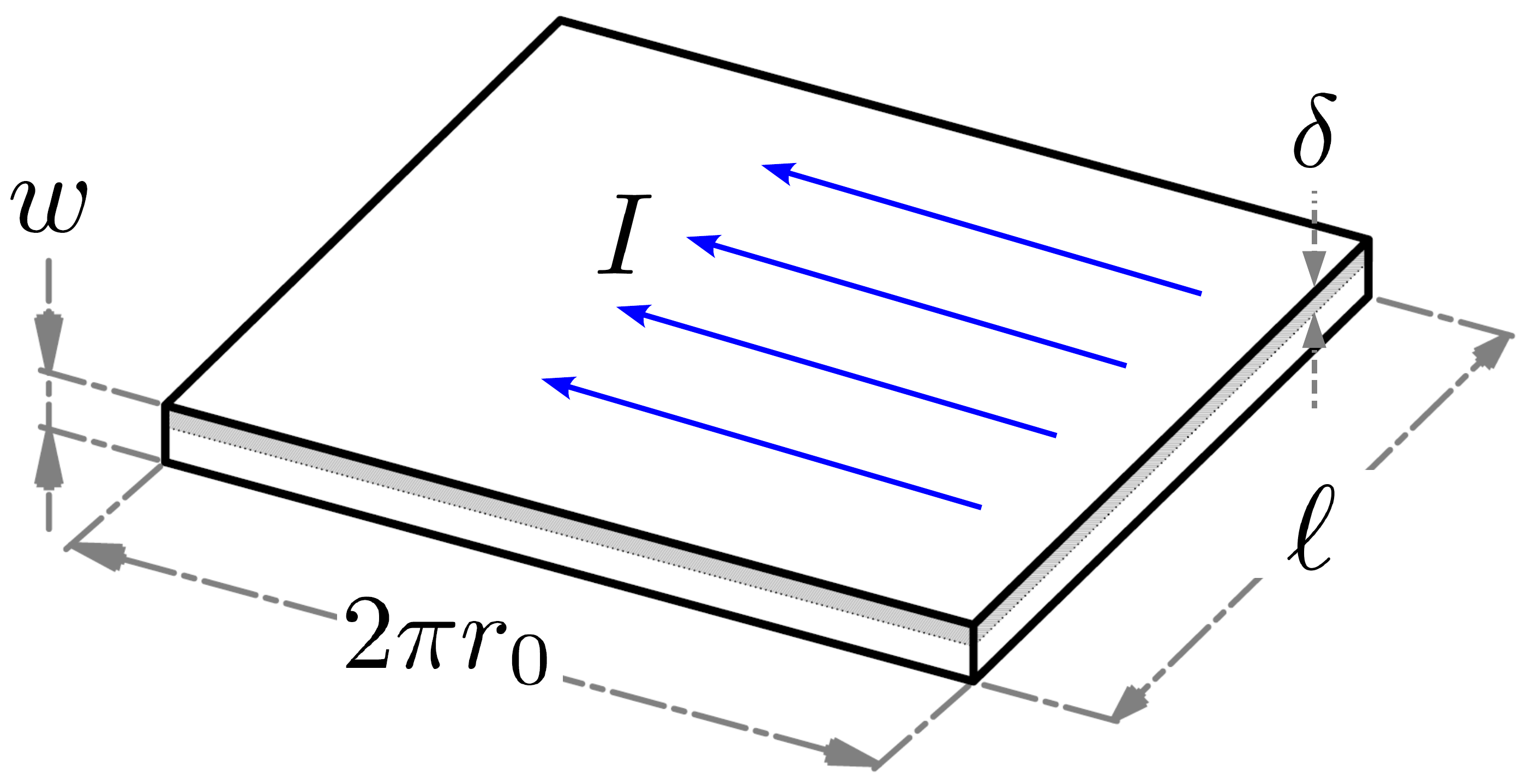} & ~~~(d)\includegraphics[width=2.4 cm]{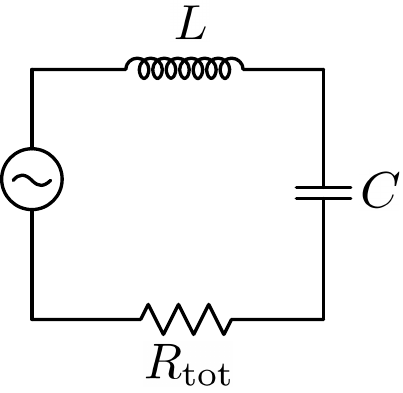}
\end{tabular}
\caption{(a) A SRR is constructed from a hollow conducting cylinder with a slit or gap along its length.  The hollow cylinder acts as a single-turn inductor, the gap acts as a capacitor, and resistance to the flow of charge is, in part, determined from the resistivity $\rho$ of the conducting material.  (b) An end view of the SRR which has an inner radius $r_0$ and a gap of height and width of $t$ and $w$ respectively.  The design dimensions of the SRR used in this study were $\ell=10.16$~cm, $r_0=1.746$~cm, $w=0.794$~cm, and $t=0.254$~mm.  (c) If a time-varying magnetic flux is applied parallel to the axis of the SRR, induced currents will be established on the inner surface of the SRR.  Due to the skin effect, the current will be confined to be within a skin depth $\delta$ of the inner surface.  Viewing the SRR is a flattened sheet shows that the current is confined to the shaded region and is directed perpendicular to the length $\ell$. (d) The SRR can be modelled very effectively as a series $LRC$ circuit.}
\label{fig:SRR}
\end{figure}

The SRR geometry is shown in Figs.~\ref{fig:SRR}(a) and (b) where the dimensions are labelled as in Ref.~\onlinecite{Hardy:1981}.  
The hollow conducting cylinder can be modelled as a single-turn solenoid with inductance:
\begin{equation}
L=\frac{\mu_0\pi r_0^2}{\ell}\label{eq:L}
\end{equation}
where $\mu_0$ is the permeability of free space and the capacitance of the gap is:
\begin{equation}
C=\varepsilon_0\frac{w\ell}{t}\label{eq:C}
\end{equation}
where $\varepsilon_0$ is the permittivity of free space and, for the moment, it is assumed that the capacitor gap is filled with air.  In Sec.~\ref{sec:III} this constraint will be relaxed.  The resonant frequency of an $LC$ resonator is given by:
\begin{equation}
\omega_0=2\pi f_0=\frac{1}{\sqrt{LC}}=\frac{c}{r_0}\sqrt{\frac{t}{\pi w}}.\label{eq:f0}
\end{equation}
Note that the SRR resonance frequency is independent of the length $\ell$.

For a series $LRC$ circuit, the quality factor $Q$ is given by:
\begin{equation}
Q=\frac{1}{R_\mathrm{tot}}\sqrt{\frac{L}{C}}=\omega_0\frac{L}{R_\mathrm{tot}}.\label{eq:Q}
\end{equation}
where, as will be seen in the following two sections, $R_\mathrm{tot}$ allows for the possibility that there are multiple mechanisms contributing to the total effective resistance in the lumped circuit element model.  When the gap of the capacitor is filled with a lossless substance (such as air), in principle, $R_\mathrm{tot}$ will be dominated by losses arising from the resistivity of the material used to construct the SRR.  If a time-varying magnetic flux is applied along the axis of the SRR, a current will be induced on its inner surface which penetrates a skin depth $\delta$ into the material.  If one imagines flattening the SRR as in Fig.~\ref{fig:SRR}(c), one effectively has a current directed through a cross-sectional area of $\delta\,\ell$ and along a length $2\pi r_0$ in a material of resistivity $\rho$ such that:
\begin{equation}
R_\delta(\omega)=\rho\frac{2\pi r_0}{\delta\,\ell}.\label{eq:R}
\end{equation}  
Combining Eqs.~\ref{eq:L}, \ref{eq:Q}, and \ref{eq:R} and using the known expression for the skin depth \mbox{$\delta=\left(2\rho/\mu_0\omega_0\right)^{1/2}$} produces the simple result:
\begin{equation}
Q=\frac{r_0}{\delta}.\label{eq:Qdelta}
\end{equation}

\begin{figure}[t] 
\centering
(a)\includegraphics[width=5.3 cm]{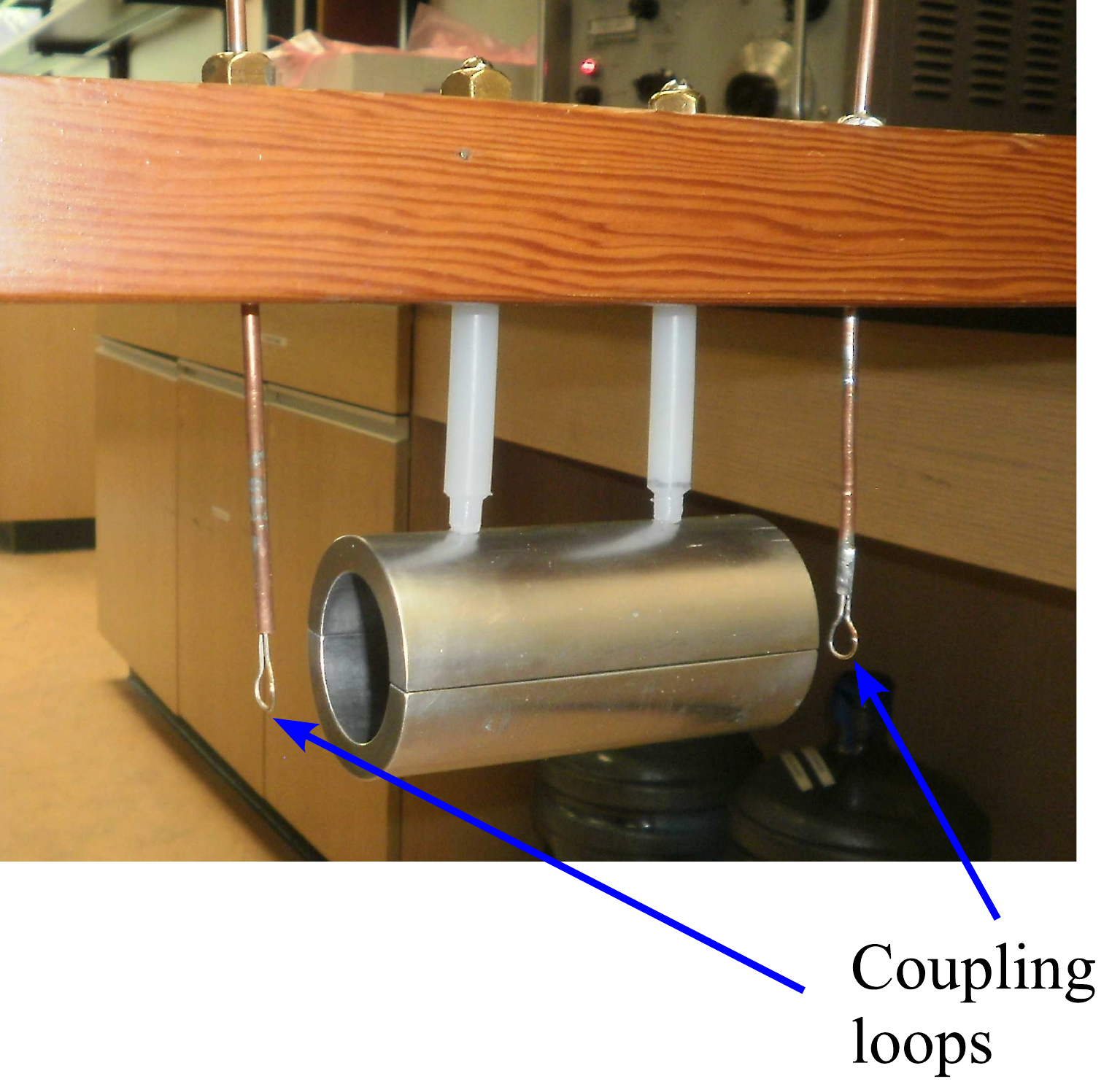}\\~\\(b)\includegraphics[width=8 cm]{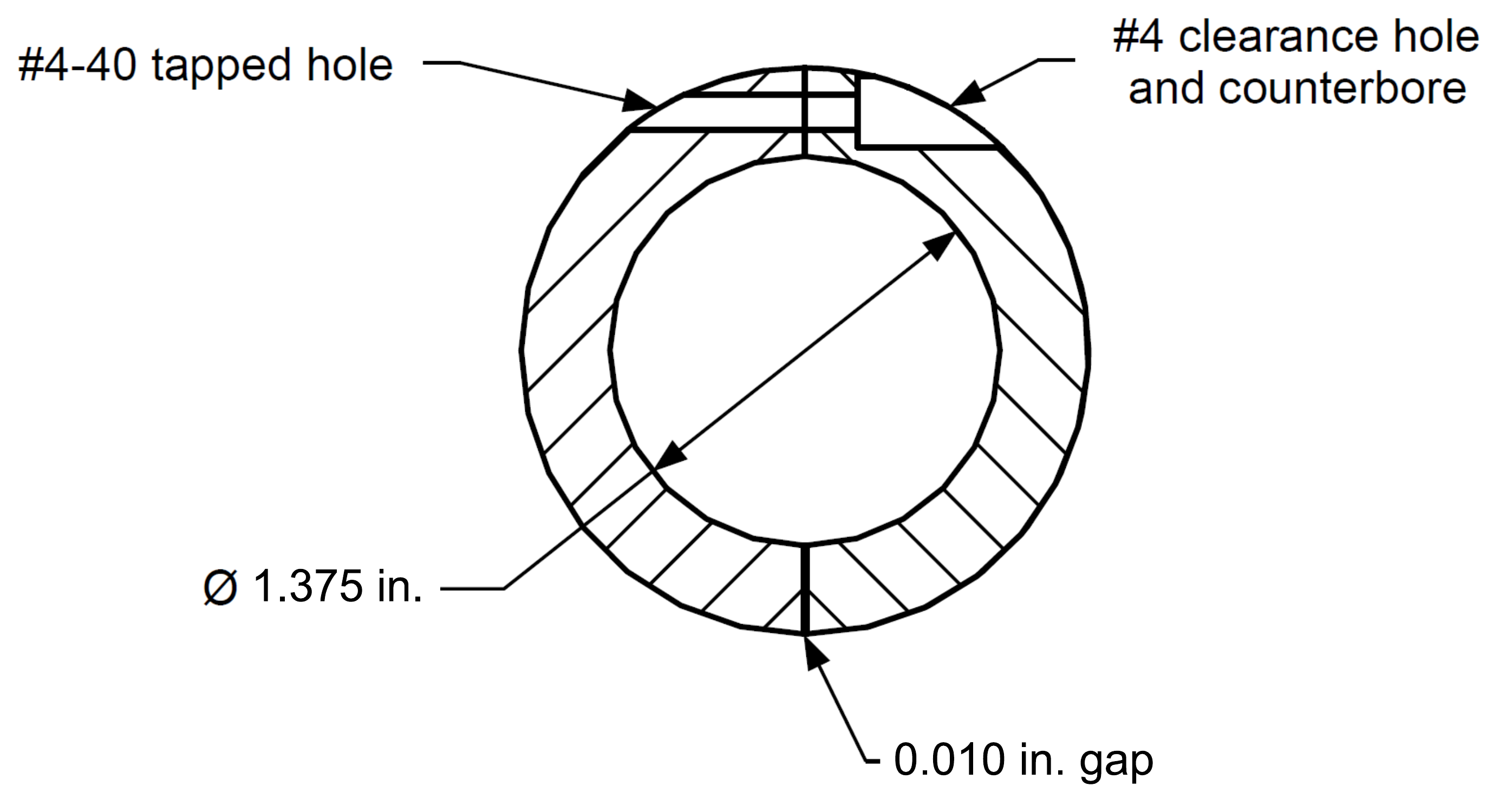}
\caption{(a) Digital photograph of the aluminum SRR suspended in air via two Teflon support rods.  Two coupling loops made using semi-rigid coaxial cable are also shown.  One drives the resonator while the other measures the response.  The coupling loops are located 1~in.\ from the ends of the SRR. (b) Scale drawing of the cross-section of the SRR used in this experiment.  The outside diameter is 2.00~in.\ and the length is 4.00~in.  To facilitate making the narrow gap ($t=0.010$~in.) needed to create an appreciable capacitance, the SRR was assembled from two half-rings that were then bolted together using \#4-40 screws as shown.  The edge of each half-ring used to form the capacitive region was machined 0.005~in.\ undersized.}
\label{fig:digital}
\end{figure}

The experimental geometry used to characterize the resonance of the SRR while suspended in air is shown in Fig.~\ref{fig:digital}.
The SRR used in this experiment was fabricated from a 2.00~in.\ diameter aluminum rod as the material was readily available in the laboratory.  Note that, because of its lower resistivity, a copper SRR would result in a slightly higher $Q$.  For a fixed outer diameter $D$, the resonance frequency of the SRR is minimized if the inner diameter is set to be $d=2D/3$.  As shown in Fig.~\ref{fig:digital}(a), coupling loops were made by shorting the center conductor of a semi-rigid 0.141~in.\ outer diameter coaxial cable to its outer conductor.  One of the coupling loops was connected to a signal generator and was used to couple EM energy into the SRR.  In this work, we used a Hewlett Packard model 608C VHF signal generator (10--480~MHz).  At the opposite end of the SRR, a second coupling loop connected to a Hewlett Packard model 8555A spectrum analyzer was used to probe the response of the SRR.

\begin{figure}[t] 
\centering
\includegraphics[width=\columnwidth]{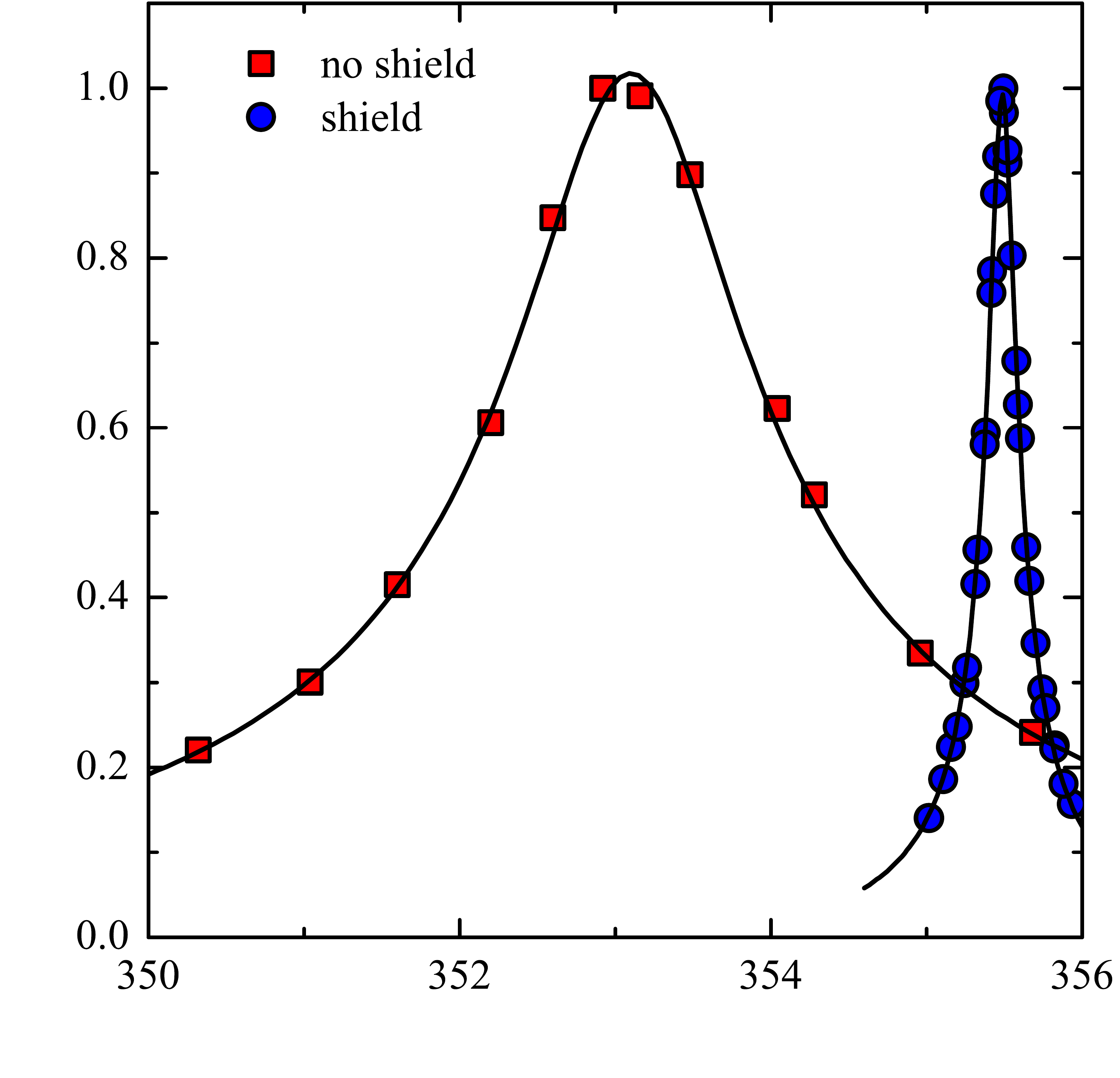}
\caption{Characterization of the resonance of the SRR when suspended in air.  Red squares: when not contained within a conducting tube, losses due to EM radiation result in a reduced quality factor (\mbox{$f_0=353.10\pm0.03$~MHz}, \mbox{$Q=244\pm 5$}).  Blue circles: When inside a conducting cylinder, radiative losses are suppressed resulting in a substantially higher quality factor.  The conducting enclosure also causes a small upwards shift in the resonance frequency (\mbox{$f_0=355.487\pm 0.005$~MHz}, \mbox{$Q=2050\pm 20$}).}
\label{fig:air}
\end{figure}

The in-air resonance of the SRR was characterized by measuring the detected signal amplitude as a function of the signal generator frequency.  The output power of the signal generator was set to $-35$~dBm such that, in the absence of the SRR, there was no detectable direct-coupling of the signal from one loop to the other.  It was also verified that the shape of the measured resonance was independent of the generator output power.  The results are shown by the red squares in Fig.~\ref{fig:air}.  
A weighted least-squares fit to a Lorentzian was performed and the best-fit line is also shown in the figure.  The typical error bar is approximately equal to the size of the data points in the plot.  The Lorentzian lineshape is given by:
\begin{equation}
V_\mathrm{N}=\frac{1}{\sqrt{1+\left(2\pi\dfrac{f}{\Gamma}\right)^2\left[1-\left(\dfrac{f_0}{f}\right)^2\right]}}
\end{equation}
where $f$ is frequency and $\Gamma=R_\mathrm{tot}/L$ characterizes the width of the resonance such that \mbox{$Q=\omega_0/\Gamma$}.  The measured in-air resonance frequency of 353~MHz is somewhat higher than the value predicted using Eq.~\ref{eq:f0} (275~MHz) due primarily to the fact that it is difficult to accurately measure the gap dimension $t$.  The measured quality factor $Q=244$ is much less than the value predicted by Eq.~\ref{eq:Qdelta} which is 2280, assuming \mbox{$\rho=8.2\times 10^{-8}~\Omega\,\mathrm{m}$} for aluminum and using the measured resonance frequency.  The reason for this large discrepancy is that, in deriving Eq.~\ref{eq:Qdelta}, radiative losses have been neglected resulting in an underestimate of the net effective resistance.  It is easy to verify that radiative effects are important, as waving one's hand in the vicinity of the SRR will significantly alter the measured resonance.

\begin{figure}[t] 
\centering
\includegraphics[width=\columnwidth]{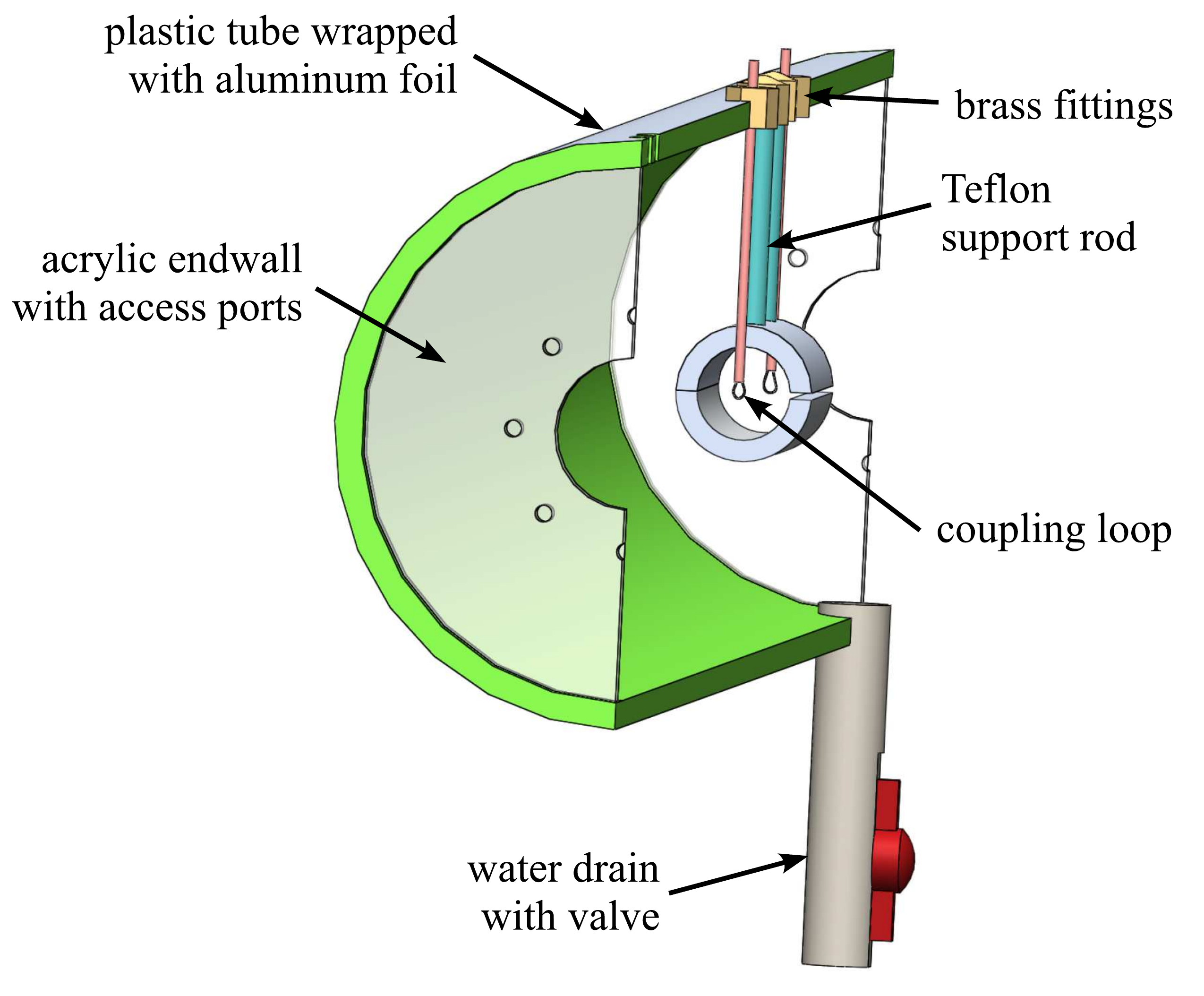}
\caption{Drawing of the SRR and coupling loops mounted inside a conducting tube (shown in cross-section).  The drawing is to scale except that the gap in the SRR has been exaggerated by a factor of five for clarity.  The EM shield is a length of plastic sewer pipe that has been wrapped with multiple layers of aluminum foil.  The ends of the waveguide are equipped with watertight endwalls that have sealable access ports.  Finally, there are openings at the top and a drain at the bottom used for filling and removing water.}
\label{fig:design}
\end{figure}

Radiative losses can be easily suppressed by surrounding the SRR by a suitable EM shield.\cite{Hardy:1981}  One convenient way to shield the resonator is by placing it in a cylindrical waveguide with the cutoff of the lowest-frequency TE$_{11}$ mode well above the resonance frequency of the SRR.  Effectively, the circumference of the waveguide is required to be less than $c/f_0$.  The geometry used in this experiment is shown in Fig.~\ref{fig:design}. 
Large metal pipes are expensive and heavy, so in this work we used a 97~cm length of a 214~mm outer diameter plastic sewer pipe that was wrapped with four or five layers of aluminum foil.  At 350~MHz the skin depth of aluminum is approximately $8~\mu$m so that effective shielding can be obtained using any practical thickness of aluminum foil. The thickness of the foil used in this work was $25~\mu$m. 

The presence of the cylindrical shield modifies both $f_0$ and $Q$ of the SRR resulting in geometrical correction factors that need to be included in Eqs.~\ref{eq:f0} and \ref{eq:Qdelta}.  These factors are derived by Hardy and Whitehead in Ref.~\onlinecite{Hardy:1981} and will not be repeated here.  It is worth noting, however, that the corrections are small provided that cross-sectional area of the SRR is much less than that of the cylindrical shield.  For example, for the dimensions used in this experiment the predicted resonance frequency and quality factor increase by only 1.4\% and 2.2\% respectively.  The blue circles in Fig.~\ref{fig:air} show the measured resonance with the cylindrical shield in place.  Clearly visible are a large increase in the $Q$ \mbox{($2050\pm 20$)} due to the suppression of radiative losses and a slight upwards shift of $f_0$ due to the geometrical correction.  The measured $Q$ increased substantially and falls only 10\% below the predicated value.  We speculate that lower $Q$ results from the fact that the SRR was constructed from two separate pieces that were then bolted together (see Fig.~\ref{fig:digital}(b)).  The electrical joint between the two halves likely results in an additional series resistance that effectively increases $R_\mathrm{tot}$ in Eq.~\ref{eq:Q}.  

\section{Immersed in water}\label{sec:III}

This section also makes use of the experimental geometry shown in Fig.~\ref{fig:design}. However, after sealing the access ports using acrylic plates and rubber o-rings, the cylindrical waveguide was filled with distilled water.  The distilled water used was measured to have a conductivity less than \mbox{$2\times 10^{-6}~\Omega^{-1}\mathrm{m}^{-1}$}.  The water fills the space between the capacitor plates of the SRR and acts as a dielectric.  At the frequencies of interest, the dielectric constant of water is large ($\sim 80$) resulting in a corresponding increase in the SRR capacitance and decrease in the resonant frequency.

To understand the effect of the water on the $Q$ of the SRR, we start by analyzing the effective impedance:
\begin{equation}
Z=\left[R_\delta(\omega)+R_\mathrm{ex}\right]+\frac{1}{j\omega\varepsilon C}+j\omega L
\end{equation}
where $L$, $C$, and $R_\delta(\omega)$ are given by Eqs.~\ref{eq:L}, \ref{eq:C}, and \ref{eq:R} respectively, $\varepsilon$ is the relative permittivity of water, and $R_\mathrm{ex}$, assumed to be frequency independent, represents any unaccounted for extrinsic losses associated with the SRR such as the effective resistance resulting from the electrical joint between the two halves of the SRR and possibly coupling effects.    Equation~\ref{eq:R} can be used to estimate \mbox{$R_\delta(\omega)\approx 11.6~\mathrm{m}\Omega$} and the measured resonance in air can be used to estimate that \mbox{$R_\delta(\omega)+R_\mathrm{ex}=\omega_0L/Q=12.9\pm 0.1~\mathrm{m}\Omega$} at 355~MHz suggesting that \mbox{$R_\mathrm{ex}\approx 1.3~\mathrm{m}\Omega$}.  In general, the relative permittivity has both real and imaginary components $\varepsilon=\varepsilon^\prime-j\varepsilon^{\prime\!\prime}$ such that the impedance given above can be re-expressed as:
\begin{align}
\begin{split}
Z&=\left[R_\delta+R_\mathrm{ex}+\frac{1}{\omega C}\left(\frac{\varepsilon^{\prime\!\prime}}{\left(\varepsilon^\prime\right)^2+\left(\varepsilon^{\prime\!\prime}\right)^2}\right)\right]\\
&\qquad{}+\frac{1}{j\omega C}\left(\frac{\varepsilon^\prime}{\left(\varepsilon^\prime\right)^2+\left(\varepsilon^{\prime\!\prime}\right)^2}\right)+j\omega L
\end{split}\\
&\approx\left[R_\delta+R_\mathrm{ex}+\frac{\varepsilon^{\prime\!\prime}}{\omega C\left(\varepsilon^\prime\right)^2}\right]+\frac{1}{j\omega\varepsilon^\prime C}+j\omega L\\
&=\left[R_\delta+R_\mathrm{ex}+R_{\varepsilon^{\prime\!\prime}}(\omega)\right]+\frac{1}{j\omega\varepsilon^\prime C}+j\omega L
\end{align}
where in second expression we have made use of the fact that, in the frequency range of interest, it is expected that $\varepsilon^\prime\gg\varepsilon^{\prime\!\prime}$.\cite{Buchner:1999}  This analysis very neatly shows that $\varepsilon^\prime$ characterizes the polarizability of the dielectric material while $\varepsilon^{\prime\!\prime}$ characterizes dissipation. Applying the standard series $LRC$ circuit results to this impedance gives:
\begin{align}
\omega_\mathrm{H_2O}&=\frac{\omega_0}{\sqrt{\varepsilon^\prime}}\\
\frac{1}{Q_\mathrm{H_2O}}&=\left[R_\delta(\omega)+R_\mathrm{ex}+R_{\varepsilon^{\prime\!\prime}}(\omega)\right]\sqrt{\frac{\varepsilon^\prime C}{L}}\\
&=\frac{1}{Q_\delta}+\frac{1}{Q_\mathrm{ex}}+\frac{1}{Q_{\varepsilon^{\prime\!\prime}}}
\end{align} 
where $\omega_\mathrm{H_2O}$ and $Q_\mathrm{H_2O}$ are the resonant frequency and quality factor when the SRR is submerged in pure water.  The ratio of $\omega_0$ to $\omega_\mathrm{H_2O}$ gives an accurate measurement $\varepsilon^\prime$ and $Q_\mathrm{H_2O}$ can be used to estimate $\varepsilon^{\prime\!\prime}$.  Note that this analysis very clearly demonstrates that the quality factors associated with different mechanisms separate and their inverses are summed to give the inverse of the net $Q$, a concept likely unfamiliar to many undergraduate students. 

\begin{figure}[t] 
\centering
\includegraphics[width=\columnwidth]{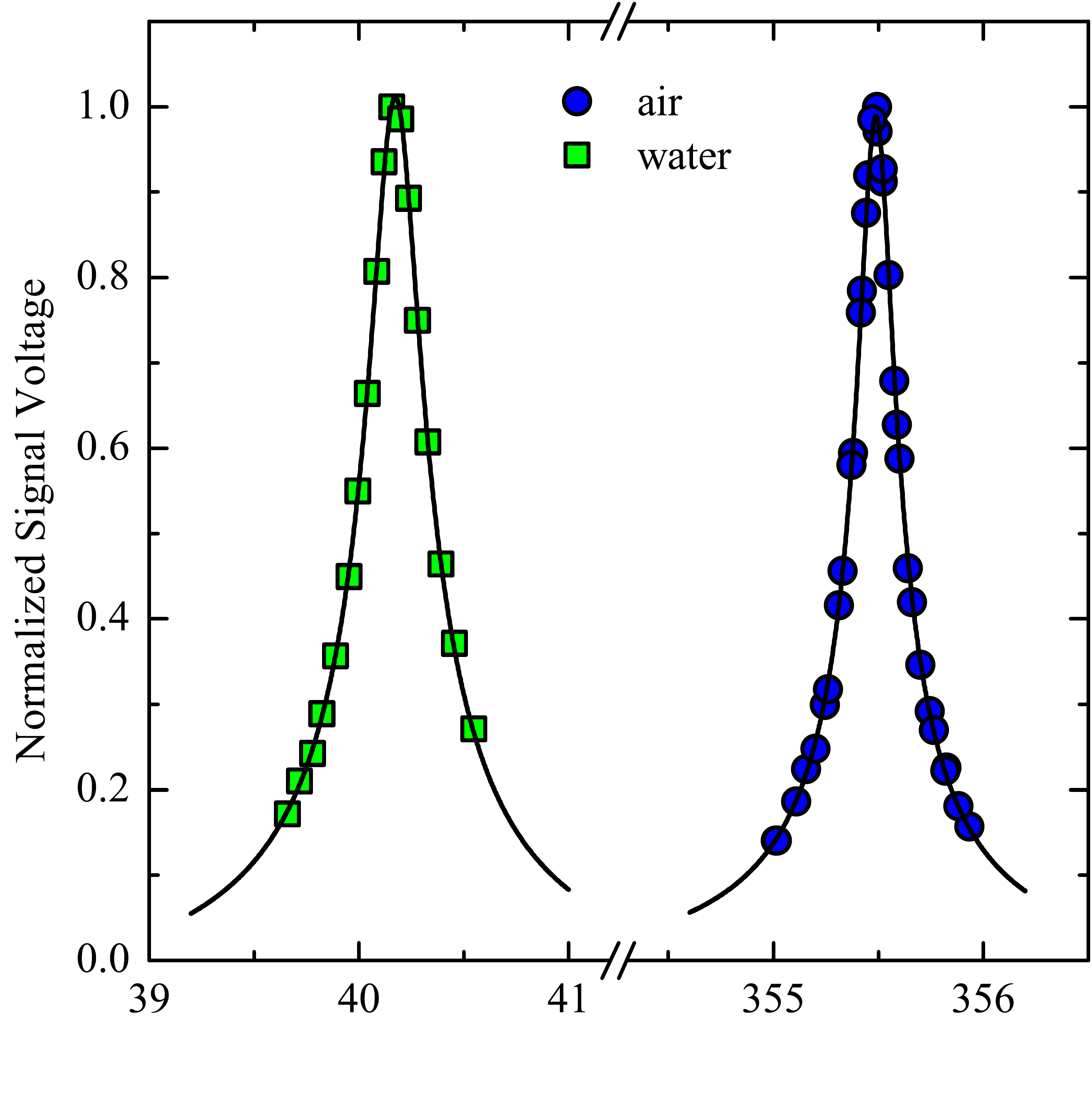}
\caption{The normalized resonance measured when the SRR was submerged in water (green squares: \mbox{$f_\mathrm{H_2O}=40.174\pm0.005$~MHz}, \mbox{$Q_\mathrm{H_2O}=159\pm 3$}) and air (blue circles: \mbox{$f_0=355.487\pm 0.005$~MHz}, \mbox{$Q=2050\pm 20$}).  Note the break in the frequency axis.  The blue circles are the same data previously shown in Fig.~\ref{fig:air}.}
\label{fig:water}
\end{figure}

The in-water resonance is shown as the green squares in Fig.~\ref{fig:water}.
From the measured data, \mbox{$\varepsilon^\prime=\left(f_0/f_\mathrm{H_2O}\right)^2=78.30\pm 0.02$} which agrees remarkably well with Buchner {\it et al.}\ who measured $\varepsilon^\prime=78.32$ at $25^\circ$C.\cite{Buchner:1999}  Equation~\ref{eq:Qdelta} was used to calculate \mbox{$Q_\delta=768$} at 40~MHz; from the estimate of $R_\mathrm{ex}$, one finds \mbox{$Q_\mathrm{ex}\approx\omega_\mathrm{H_2O}L/R_\mathrm{ex}=2300$}; and from the green squares in Fig.~\ref{fig:water}, $Q_\mathrm{H_2O}$ was measured to be 159.  Taken together, these results can be used to determine $Q_{\varepsilon^{\prime\!\prime}}=220$ and \mbox{$R_{\varepsilon^{\prime\!\prime}}=\omega_\mathrm{H_2O}L/Q_{\varepsilon^{\prime\!\prime}}=13.6~\mathrm{m}\Omega$}. It is important to note that all of these calculations have been done in terms of $\omega_\mathrm{H_2O}$ and $L$ rather than $C$ since the dimension $t$ is not known accurately. 

Finally, the imaginary part of the relative permittivity can be calculated from \mbox{$\varepsilon^{\prime\!\prime}=R_{\varepsilon^{\prime\!\prime}}\varepsilon^\prime/\left(\omega_\mathrm{H_2O}L\right)=0.36$} which does not agree with an extrapolation of the results of Buchner and co-workers which gives $\varepsilon^{\prime\!\prime}=0.15$ at 40~MHz.\cite{Buchner:1999}  There are numerous possibilities for this discrepancy: the Buchner {\it et al}.\ measurements did not extend below 200~MHz requiring an extrapolation to 40~MHz which may not be reliable, the extrinsic losses characterized by $R_\mathrm{ex}$ were assumed to be frequency independent which may not be valid, and the assumed value for the resistivity of aluminum may be slightly off for the material used to construct the SRR.  Nevertheless, the measurement clearly demonstrates that the permittivity of water is complex and that the SRR quality factor is sensitive to the imaginary component.  The quality of this measurement could be improved by designing a SRR that operates at a higher frequency where $\varepsilon^{\prime\!\prime}$ is larger and by building it out of copper to enhance $Q_\delta$.

Here, one practical point is noted before concluding the section with a brief discussion of the use of SRRs in current research. Immediately after filling the cylindrical waveguide with water, SRR resonators with very small gaps will likely trap many small air bubbles which will make it impossible to accurately measure $\varepsilon^\prime$.  These air bubbles are visible by eye and one way to remove them is by holding the top of the support rods and gently swaying the SRR back a forth while completely submerged in water.  For the SRR used in this project, it was found this procedure successfully removed all of the visible air bubbles within two to three minutes and the subsequent measurements were reliable and reproducible.

In a way completely analogous to the methods discussed in this section, researchers use microwave resonators (SRRs, cylindrical cavities and tunnel diode oscillators) to study the electrodynamic properties of superconductors.\cite{Bonn:2007, Bonn:1993, Bonn:1996, Kokales:2000, Peligrad:1998, Broun:1997, Prozorov:2006}  The resonance frequency probes the so-called penetration depth, which characterizes how far into the surface of a superconductor external magnetic fields penetrate, and the quality factor can be used to determine the surface resistance of the superconducting sample.  A key difference is that, whereas in the current experiment water in the gap of the SRR interacts with the electric field and modifies the effective capacitance, in measurements of superconductors the sample is placed in a region of high magnetic field.  In the case of a SRR, a superconducting sample is placed in the bore of the resonator.  When cooled below the superconducting transition temperature, due to its diamagnetic response (relative permeability $\mu_\mathrm{r}=0$), the sample expels the magnetic flux from its interior to within a temperature-dependent penetration depth $\lambda(T)$ of its surface.\cite{Tinkham:1996}  The change in magnetic flux within the bore of the resonator results in a modification of the SRR inductance and therefore the resonant frequency $f_0$.  In practice, the sample temperature $T$ is typically controlled independently of the SRR and one tracks changes in resonant frequency as a function of $T$ to deduce the temperature dependence of the penetration depth.\cite{Hardy:1993, Bobowski:2010}  Another point of interest is that the surfaces of microwave resonators used for high-precision measurements are typically plated with a thin superconducting layer (usually a lead-tin alloy) and cooled to low temperatures so as to minimize losses associated with the SRR itself.  Quality factors of unloaded SRRs can exceed $10^6$ leading to frequency resolutions that are better than one part in $10^9$.\cite{Hardy:1993, Bobowski:2010}

\section{Immersed in aqueous salt solution}\label{sec:IV}
\begin{figure}[t] 
\centering
\includegraphics[width=\columnwidth]{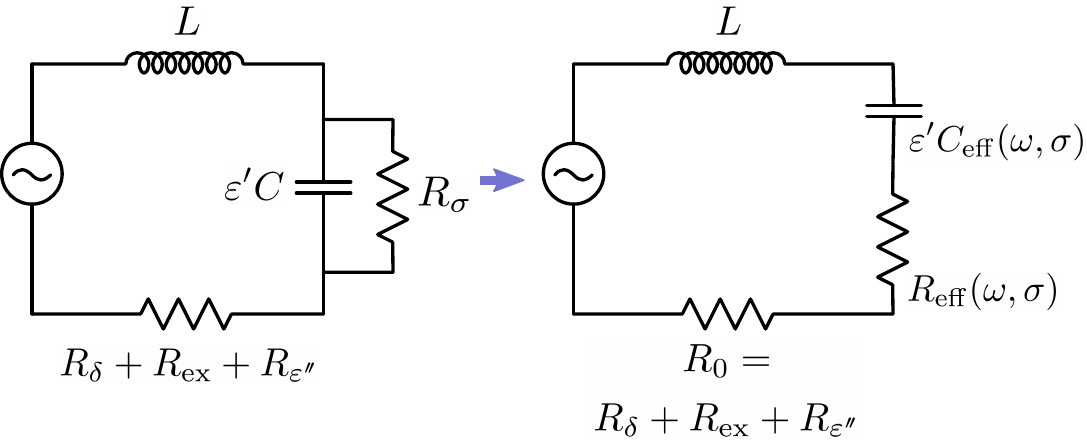}
\caption{The left-hand figure shows the modified lumped-element circuit model resulting from submerging the SRR in an aqueous solution of NaCl.  The dissolved ions provide a parallel path for current to cross the gap of the SRR.  The parallel combination of $R_\sigma$ and $\varepsilon^\prime C$ can be re-expressed as an equivalent series combination (right-hand figure) which is more convenient as it allows for a standard series $LRC$ circuit analysis.}
\label{fig:LRCmod}
\end{figure}

The final measurement that will be discussed is an investigation into the effect of adding salt to the water surrounding the SRR.  The salt selected was table salt obtained from a local supermarket and was 99.7\% NaCl.  When dissolved in water, the dissociated Na$^+$ and Cl$^{-}$ ions make the aqueous solution conducting.  The ions within the gap region of the SRR are exposed to an electric field resulting in a second parallel mechanism for current conduction across the gap.  The gap resistance is given by:
\begin{equation}
R_\sigma=\frac{1}{\sigma}\frac{t}{w\ell}
\end{equation}
where $\sigma$ is the conductivity of the aqueous NaCl solution.  The lumped-element circuit model is shown on the left-hand side of Fig.~\ref{fig:LRCmod}.  
It is convenient, and always possible, to rewrite the impedance of the parallel combination of $R_\sigma$ and $\varepsilon^\prime C$ as an equivalent series combination of $R_\mathrm{eff}(\omega, \sigma)$ and $\varepsilon^\prime C_\mathrm{eff}(\omega, \sigma)$ as follows:\cite{Horowitz:1980}
\begin{align}
\begin{split}
R_\sigma\parallel Z_{\varepsilon^\prime C} &=\frac{R_\sigma}{1+\left(\omega\varepsilon^\prime R_\sigma C\right)^2}\\
&\qquad{}+\frac{1}{j\omega\varepsilon^\prime C}\left[\frac{\left(\omega\varepsilon^\prime R_\sigma C\right)^2}{1+\left(\omega\varepsilon^\prime R_\sigma C\right)^2}\right]
\end{split}\\
&=R_\mathrm{eff}(\omega,\sigma)+\frac{1}{j\omega\varepsilon^\prime C_\mathrm{eff}(\omega,\sigma)}
\end{align}
where $Z_{\varepsilon^\prime C}=\left(j\omega\varepsilon^\prime C\right)^{-1}$ is the impedance of capacitor $C$ filled with a dielectric of relative permittivity $\varepsilon^\prime$.
Note that the time constant $\varepsilon^\prime R_\sigma C=\varepsilon^\prime\varepsilon_0/\sigma$ is independent of the SRR dimensions.  In the low-conductivity limit, such that \mbox{$\omega\varepsilon^\prime\varepsilon_0/\sigma\gg 1$}, $C_\mathrm{eff}\approx C$ and the total impedance of the equivalent series circuit becomes:
\begin{align}
Z&\approx\left[R_0+\frac{\sigma}{\left(\omega \varepsilon^\prime\varepsilon_0\right)^2}\frac{t}{w\ell}\right]+\frac{1}{j\omega\varepsilon^\prime C}+j\omega L\label{eq:approx1}\\
&=\left[R_0+R_\mathrm{eff}(\omega,\sigma)\right]+\frac{1}{j\omega\varepsilon^\prime C}+j\omega L\label{eq:approx2}
\end{align} 
where \mbox{$R_0=R_\delta+R_\mathrm{ex}+R_{\varepsilon^{\prime\!\prime}}$} represents the total resistance when $\sigma=0$.  Based on the results of the previous section, the low-conductivity limit will be satisfied provided \mbox{$\sigma\ll 0.2~\Omega^{-1}\mathrm{m}^{-1}$}.  This condition is satisfied for all of the data presented in this section such that Eqs.~\ref{eq:approx1} and \ref{eq:approx2} are always valid.  Notice the potentially counterintuitive result $R_\mathrm{eff}\propto\sigma$.  This low-conductivity expression for the impedance shows that the resonance frequency is independent of $\sigma$ and given by \mbox{$\omega_\sigma=\omega_\mathrm{H_2O}=1/\sqrt{\varepsilon^\prime C L}$} while the net quality factor is determined from:
\begin{equation}
\dfrac{1}{Q_\mathrm{net}}=\dfrac{R_0}{\omega_\sigma L}+\dfrac{\sigma}{\omega_\sigma\varepsilon^\prime\varepsilon_0}=\frac{1}{Q_\mathrm{H_2O}}+\dfrac{1}{Q_\sigma}.\label{eq:Qnet}
\end{equation}
For very low values of conductivity \mbox{($\sigma\ll\varepsilon^\prime\varepsilon_0 R_0/L\approx 1\times 10^{-3}~\Omega^{-1}\mathrm{m}^{-1}$)}, such that the losses are dominated by the $R_0$ term, one recovers the expected result $Q_\mathrm{net}\approx Q_\mathrm{H_2O}$. On the other hand, when the losses are dominated by the conductivity of the aqueous salt solution \mbox{($\varepsilon^\prime\varepsilon_0 R_0/L\ll\sigma\ll\omega\varepsilon^\prime\varepsilon_0$)}, $Q_\mathrm{net}\propto\sigma^{-1}$.

Sodium chloride was added by extracting a small amount of water ($\sim 500$~mL) from the cylindrical waveguide into which a known mass of the NaCl was fully dissolved.  This aqueous solution was then added back into the waveguide structure.  In order to observe full behaviour of $Q_\mathrm{net}$ given by Eq.~\ref{eq:Qnet}, one must start with very small concentrations of NaCl such that \mbox{$\sigma\ll\varepsilon^\prime\varepsilon_0 R_0/L$}.  In this project, the mass of NaCl was measured using a 0.01~g resolution digital scale.  If such a scale is unavailable, one could dissolve a larger mass of NaCl into a large volume of water and add it into the waveguide structure a little at a time.  Once added, sufficient time must be allotted to allow the NaCl to reach an equilibrium distribution before starting measurements.  In present the work, a minimum settling time of one hour was used.  The resonance should be actively monitored (using a spectrum analyzer, for example) and measurements started only after changes to the lineshape have subsided.      

The concentrations of NaCl used and the associated conductivities are summarized in Table~\ref{tab:sigma}. 
\begin{table}[htb]
\centering
\begin{ruledtabular}
\begin{tabular}{c c c c c}
mass & $N$  & $\sigma$ & $Q_\sigma$ & $Q_\mathrm{net}$\\
(g) & $\left(10^{-4}~\mathrm{mol/L}\right)$ & $\left(10^{-3}~\Omega^{-1}\mathrm{m}^{-1}\right)$ & ~ & ~\\
\hline
0 & N/A & $<2\times 10^{-3}$ & $>87,500$ & $159\pm 3$\\
0.005 & $3.04\times 10^{-2}$ & $3.16\times 10^{-2}$ & 5530 & $180\pm 3$\\
0.010 & $6.08\times 10^{-2}$ & $6.33\times 10^{-2}$ & 2760 & $172\pm 3$\\
0.020 & 0.121 & 0.125 & 1400 & $158\pm 2$\\
0.030 & 0.181 & 0.189 & 927 & $134\pm 2$\\
0.040 & 0.242 & 0.252 & 694 & $110\pm 2$\\
0.060 & 0.364 & 0.378 & 462 & $87\pm 1$\\
0.100 & 0.608 & 0.632 & 277 & $79\pm 1$\\
0.150 & 0.912 & 0.948 & 184 & $69\pm 1$\\
0.200 & 1.22 & 1.26 & 138 & $55\pm 1$\\
0.400 & 2.43 & 2.53 & 69 & $53\pm 1$\\
0.600 & 3.65 & 3.80 & 46 & $25.2\pm 0.5$\\
0.800 & 4.87 & 5.06 & 35 & $20.0\pm 0.5$\\
1.20 & 7.31 & 7.59 & 23 & $18.9\pm 0.4$\\
1.60 & 9.74 & 10.1 & 17 & $12.0\pm 0.3$\\
2.00 & 12.2 & 12.7 & 14 & $12.9\pm 0.3$\\
2.50 & 15.2 & 15.8 & 11 & $10.0\pm 0.3$\\
3.50 & 21.3 & 22.1 & 7.9 & $8.4\pm 0.3$\\
5.00 & 30.4 & 31.6 & 5.5 & $7.8\pm0.3$
\end{tabular}
\end{ruledtabular}
\caption{Table of data corresponding to Fig.~\ref{fig:conductivity}(b) showing the mass of NaCl used, the corresponding molarity, the calculated conductivity, $Q_\sigma$ calculated using Eq.~\ref{eq:Qnet}, and the measured $Q_\mathrm{net}$.}\label{tab:sigma}
\end{table}
The molarity $N$ of the aqueous NaCl solution was calculated using the known volume of the cylindrical waveguide ($28.1\pm 1.2$~L).  In Ref.~\onlinecite{Stogryn:1971}, Stogryn provides an empirical formula that can be used to convert the molarity of NaCl to conductivity for a solution at $25^\circ$C.  For the concentrations used in this experiment, to within an excellent approximation, $\sigma=10.394 N$ where $\sigma$ has units of $\Omega^{-1}\mathrm{m}^{-1}$ and $N$ is in mol/L.

\begin{figure}[t] 
\centering
(a)\includegraphics[width=0.85\columnwidth]{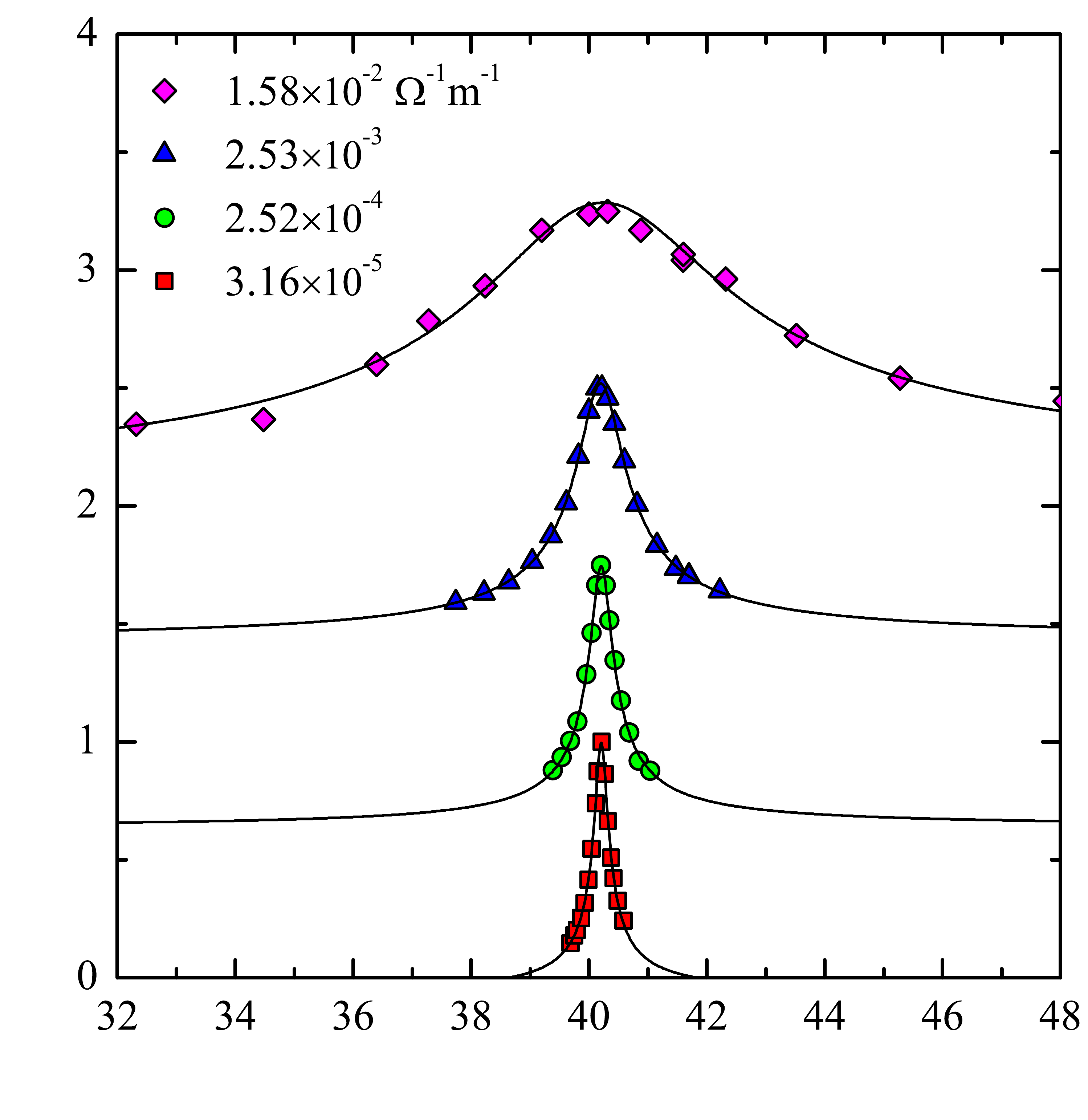}\\~\\~\\(b)\includegraphics[width=0.9\columnwidth]{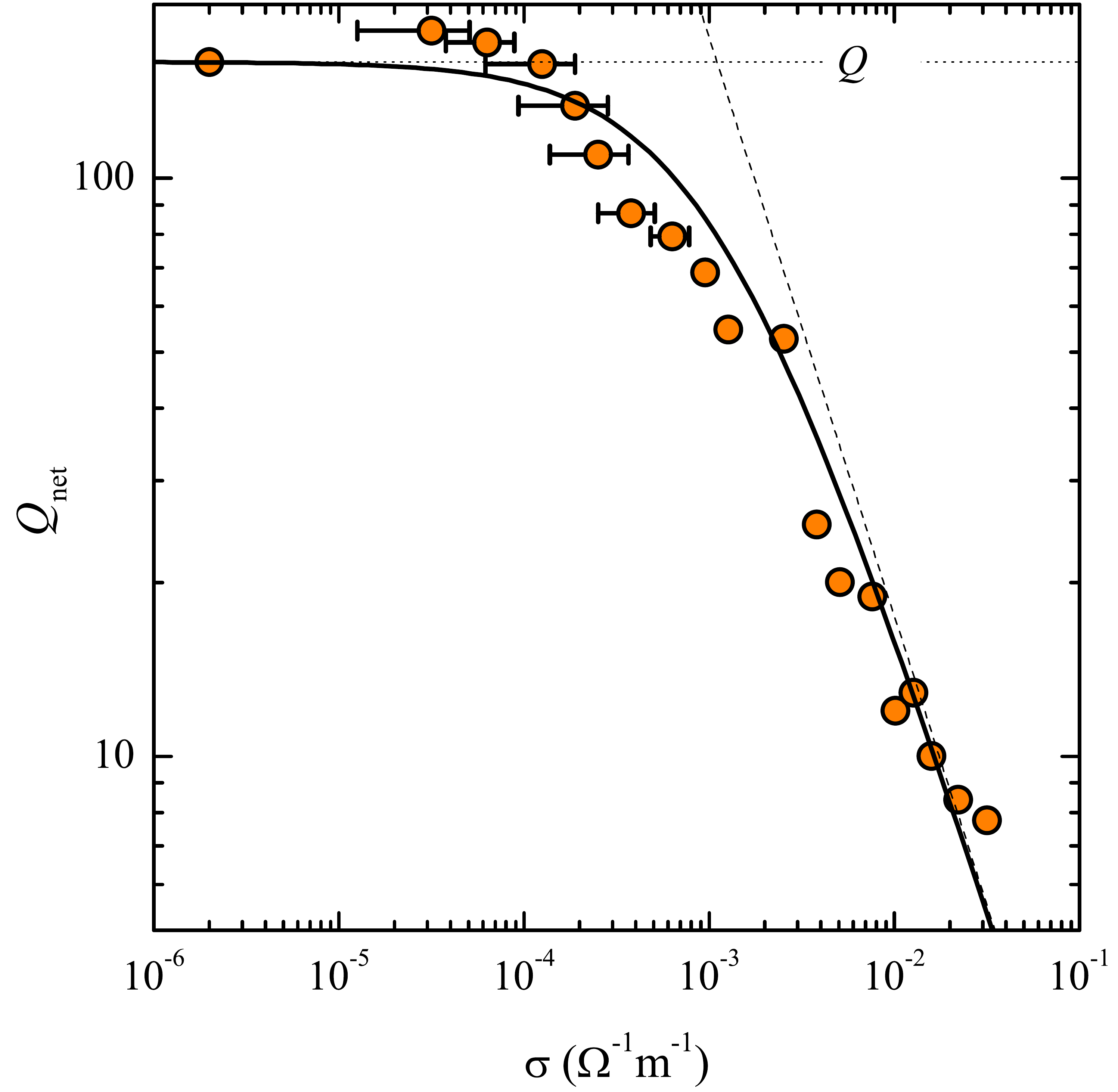}
\caption{(a) A sample of measured resonance curves and the corresponding fits for four conductivities that span $3\times 10^{-5}$ to $1.6\times 10^{-2}~\Omega^{-1}\mathrm{m}^{-1}$.  Adjacent datasets have been offset by 0.75 along the vertical axis for clarity.  An initially sharp resonance is seen to dramatically broaden as $\sigma$ is increased.  
The signal voltage error bars are approximately equal to the size of the data points. (b) Plot of $Q_\mathrm{net}$ as a function of $\sigma$ on a log-log scale.  The thick line is a prediction based on Eq.~\ref{eq:Qnet} and the known values of $Q_\mathrm{H_2O}=159$ and $\varepsilon^\prime=78.30$ from Sec.~\ref{sec:III}.  The dotted line indicates the $\sigma=0$ value of the quality factor and the dashed line follows a $\sigma^{-1}$ dependence.}
\label{fig:conductivity}
\end{figure}

Figure~\ref{fig:conductivity}(a) shows four of the 19 resonance curves measured from very low \mbox{($\sigma\ll\varepsilon^\prime\varepsilon_0 R_0/L$)} to relatively high \mbox{($\varepsilon^\prime\varepsilon_0 R_0/L\ll\sigma\ll\omega\varepsilon^\prime\varepsilon_0$)} NaCl concentrations.
As predicted, there is no discernible change in the resonance frequency as the conductivity of the water is increased.  However, for sufficiently high conductivities, the width of the resonance increases rapidly.  At all NaCl concentrations, the measured resonance fits the Lorentzian lineshape very well.  A log-log plot of the measured quality factor as a function of the conductivity is shown in Fig.~\ref{fig:conductivity}(b).  The uncertainties in $Q_\mathrm{net}$ are listed in Table~\ref{tab:sigma} and are approximately equal to the size of the data points used in plot.  For the lowest concentrations of NaCl used, the large uncertainty associated with measuring small masses of NaCl leads to a large uncertainty in the conductivity as shown by the errors bars in Fig.~\ref{fig:conductivity}(b).  As the concentration of NaCl is increased, the conductivity error bars become equal to or less than the size of the data points. The solid line shows the predicted dependence of $Q_\mathrm{net}$ on $\sigma$ given by Eq.~\ref{eq:Qnet} using the previously measured values of $Q_\mathrm{H_2O}=159$ and $\varepsilon^\prime=78.30$.  For \mbox{$\sigma\ll\varepsilon^\prime\varepsilon_0 R_0/L$}, $Q_\mathrm{net}$ plateaus as its value is determined almost entirely by $Q_\mathrm{H_20}$ which is independent of $\sigma$.  As the NaCl concentration is increased and the conductivity enters the regime \mbox{$\varepsilon^\prime\varepsilon_0 R_0/L\ll\sigma\ll\omega\varepsilon^\prime\varepsilon_0$}, $Q_\sigma$ dominates the behaviour of $Q_\mathrm{net}$ resulting in the observed $\sigma^{-1}$ dependence.  Although the solid line and data exhibit the same qualitative behaviour, there are significant differences for some values of $\sigma$.  We speculate that, despite the settling time that was allowed, these deviations may result from a slight inhomogeneity in the distribution of the dissolved NaCl.  It is possible that a small motorized turbine stirrer could be used to increase the uniformity of the aqueous solution.

\section{Summary}\label{sec:V}

To summarize, a SRR experiment suitable for physics undergraduates has been described.  The SRR resonant frequency and quality factor were predicted using a simple lumped-element circuit model.  After shielding the SRR in a suitable cylindrical waveguide, the in-air characteristics matched the predictions reasonably well.  As a research tool, SRR are used to measure the EM properties of materials.  By submerging our SRR in water the low-frequency permittivity of water was accurately and precisely determined to be \mbox{$\varepsilon^\prime=78.30\pm 0.02$}.  Enhanced broadening of the resonance implied an additional loss mechanism introduced by the water.  Analysis led to an estimate of the imaginary component of the relative permittivity of water at 40~MHz.  Designing a high-$Q$ SRR that operates at higher frequencies where the ratio $\varepsilon^{\prime\!\prime}/\varepsilon^\prime$ is larger would improve this measurement.  Lastly, the conductivity of the water was tuned by adding controlled amounts of NaCl.  For the low concentrations used, the SRR resonance frequency was predicted and subsequently verified to be independent of $\sigma$.  On the other hand, $Q$ is independent of $\sigma$ only for very low concentrations of NaCl and decreases as $\sigma^{-1}$ at higher conductivities.  The measured dependence of $Q$ on $\sigma$ agreed very well with theoretical prediction.

Some of the advantages of the current experiment are as follows:
\begin{itemize}
\item Although rf/microwave techniques are used frequently in research, the methods are not typically explored in an undergraduate setting. (An electron spin resonance-type experiment may be one of the only common exceptions.)
\item The required rf equipment is readily available or inexpensive to acquire.  For example, the generator used in this experiment was manufactured in 1954.  Additionally, the signal detector can be any power/voltage detector that operates at the required frequencies (spectrum analyzer, oscilloscope, vector network analyzer, crystal detector).
\item The electromagnetic skin depth, complex permittivity, $LRC$ resonators, complex analysis, and waveguides, are all part of a standard undergraduate physics curriculum and this experiment invites students to explore these concepts in a laboratory setting. 
\item The data collection and data analysis techniques will challenge, but not overwhelm, students.
\end{itemize}

Finally, we point out that the experiment can be easily tailored to fit the needs (or constraints) of a particular course.  At the Okanagan campus of the University of British Columbia the SRR resonator experiment is offered as an option to third- and fourth-year students taking a course in experimental physics.  Students are typically given five or six weeks to complete a project such that a motivated student would be able to reproduce most, if not all, of the work presented above.  If time is more restricted, one could design a suitable experiment around the material presented in sections~\ref{sec:II} and \ref{sec:III}.

A supplemental measurement to observe cutoff frequencies in cylindrical waveguides can be easily executed.  The measurement uses the experimental geometry shown in Fig.~\ref{fig:design} except with the SRR and its support rods removed leaving just the two coupling loops suspended inside the waveguide.  A signal launched inside the waveguide using one of the coupling loops is then be detected using the opposite coupling loop.  For frequencies below cutoff, little to no signal reaches the sensing coupling loop.  Above the cutoff frequency of the dominant TE$_{11}$ mode, the detected signal power rises sharply.  As frequency is increased further, the signal power varies in a complicated manner as higher-order waveguide modes begin to propagate.

The experiment could also easily be extended to form a more in depth undergraduate research project.  For example, placing a high-permeability sample inside the bore of the SRR would increase the inductance resulting in a decrease of the resonance frequency.  Additionally, losses associated with the sample would broaden the resonance.  However, using an iron bar sample would not work because, due to the skin effect, the magnetic field will penetrate a only a very short distance into the material.  One intriguing possibility would be to use a suspension of micron-sized iron particles as a sample.  Low-frequency relative permeabilities in excess of five have been reported for high volume fraction suspensions.\cite{deVicente:2002}  Relatively high permeability values for iron powder samples at microwave frequencies have also been reported.\cite{Wang:2004}      

\appendix*

\section{Parts and Suppliers}

This appendix provides a list of the parts needed to assemble the SRR experiment.  Where appropriate, vendors and cost estimates are also provided.

{\it Signal Generator} --  The most costly piece of equipment required is the rf signal generator.  To reproduce the SRR experiment described here requires a signal generator that operates from approximately 10 to 500~MHz.  The most affordable signal generator from Agilent Technologies \mbox{(\url{http://www.home.agilent.com})} that is suitable is the N9310A RF Signal Generator (9~kHz -- 3~GHz) which costs \$7,618.  Although we do not have experience using this instrument, the least expensive option for a new signal generator that we have found is the SG200 DDS RF Signal Generator (9~kHz -- 450~MHz) from Digimess Instruments \mbox{(\url{http://www.digimessinstruments.co.uk})} which costs approximately \$1,400.  Fortunately, many university or college physics departments will already have a suitable signal generator.  Alternatively, many dealers in used test equipment will offer calibrated signal generators at a fraction of the new purchase price.

{\it RF Detector} -- There are many options for detecting the signal transmitted to the sensing coupling loop.  These include a spectrum analyzer, a high-frequency oscilloscope, rf power meter, or a crystal detector.  If purchasing a new detector, the Agilent Technologies 423B (10~MHz -- 12.4~GHz) crystal detector retails for \$739.

{\it Coupling Loops} -- The coupling loops were fabricated from 0.141~in.\ outer diameter semi-rigid coaxial cable.  A 12~in.\ length of RG402 cable with SMA connectors can be purchased from Pasternack Enterprises \mbox{(\url{http://www.pasternack.com})} for a cost of \$38.  If cut into two 6~in.\ pieces, two coupling loops equipped with rf connectors can be made. Various flexible coaxial cables and rf adaptors that may be required can also be purchased from Pasternack Enterprises.

{\it Metal Round Bar} -- The material required to fabricate the SRR can be purchased from Metal Supermarkets \mbox{(\url{https://www.metalsupermarkets.com})}.  A 10~in.\ length of 2~in.\ diameter 6061 aluminum round bar costs \$27.  The same size bar of 110 copper costs \$189.

{\it Sewer Pipe} -- Sewer pipe can be purchased from a local supplier of plumbing equipment.  Standard sewer pipers are typically sold in  4~m lengths.  A piece of 8~in.\ diameter sewer pipe retails for approximately \$81.

{\it Acrylic Plate} -- The ends of the sewer pipe were equipped with sealable acrylic plates.  Industrial plastic suppliers sell 3/8~in.\ thick acrylic sheets for \$12 per square foot.

{\it Miscellaneous} -- The following items are all inexpensive and can be found at most hardware stores: J-B MarineWeld epoxy to create a watertight seal between the sewer pipe and the acrylic end plates, plastic (or Teflon) rods used to suspend the SRR (about 1~cm in diameter), 3/8~in.\ national pipe thread (NPT) brass plugs used to feed the coupling loops and SRR support rods through the sewer pipe, 3/4~in.\ NPT PVC ball valve, heavy duty aluminum foil, and packing tape to wrap around the last layer of aluminum foil to prevent tearing.

\begin{acknowledgments}

The author benefited from numerous enlightening discussions with Walter Hardy and Thomas Johnson during the development of this experiment.  The assistance and availability of the UBC Okanagan machine shop is also gratefully acknowledged.

\end{acknowledgments}

\end{document}